\documentstyle[amstex,amssymb]{fa98}
\Year{1998}

\newcommand{\M}{{\Bbb{M}}}
\newcommand{\half}{\textstyle{\frac12}}
\newcommand{\boldsigma}{\mbox{\boldmath$\sigma$}}
\newcommand{\boldcdot}{\mbox{\boldmath$\cdot$}}
\newcommand{\supp}{{\mathrm{supp}}}
\newcommand{\ess}{{\mathrm{ess}}}

\begin{document}
\Title{The Dirac equation without spinors}
\Shorttitle{The Dirac equation without spinors}
\By{{\sc Daniel Elton}, {\sc Dmitri Vassiliev}}
\Names{Elton, Vassiliev}
\Email{D.Vassiliev@sussex.ac.uk}
\maketitle
\begin{abstract}
In the first part of the paper we give a tensor version of the
Dirac equation. In the second part we formulate
and analyse a simple model equation which for weak external
fields appears to have properties similar to those
of the 2--dimensional Dirac equation.
\end{abstract}

\newsection{Introduction}

The paper is an attempt at analysing the geometrical meaning
of the Dirac equation
\begin{equation}\label{Dirac}
\gamma^\varkappa(i\nabla-e A)_\varkappa\psi=m\psi,
\end{equation}
and at presenting this geometry in a mathematical language
understandable to specialists in partial differential equations.
The paper consists of two parts.

In the first part we give a tensor interpretation of bispinors
(Section 3) and reformulate the Dirac equation
in tensor form (Section 4).
In Section 5 we discuss the simplifications which
occur when we move from $\M^4$ (3--dimensional Dirac equation)
to $\M^3$ (2--dimensional Dirac equation).
Note that the fact that the Dirac
equation can, in principle, be written in tensor form is not
new; see, for example, \cite{BT}.

In the second part we adopt the point of view
that the Dirac equation is not a precise equation, but
a first approximation in an asymptotic process described
by Feynman diagrams, and it should be used only when
the external field $A$ is weak, smooth, and slowly varying.
We suggest
the following model equation in $\M^3$:
\begin{equation}\label{model}
\operatorname{curl}_A u=\pm m u\,.
\end{equation}
Here $u$ is a complex valued vector function, and
$\operatorname{curl}_A\,$ is the usual $\operatorname{curl}\,$
corrected by the electromagnetic field; see next section for
precise definition.
We show that (\ref{model}) has properties very similar to those
of the two--dimensional Dirac equation.
In Section 6 we give a mathematical meaning to
``weak, smooth, and slowly varying'' by
introducing asymptotic scaling in terms of a small
dimensionless parameter $\alpha$ (fine structure constant)
and perform a formal asymptotic analysis of (\ref{model}).
This formal asymptotic analysis reduces (\ref{model}) to
the Pauli equation.
Mathematically rigorous (but, inevitably, more restrictive)
results are presented in Section 7.

Finally, in Section 8 we point out the remarkable similarity
between (\ref{model}) and the Maxwell equations.

\newsection{Principal notation}

Our notation mostly follows \cite{LL4}, with minor modifications.

By $\M^n$ we denote Minkowski $n$--space
with temporal coordinate $x^0$,
spatial coordinates $(x^1,\ldots,x^{n-1})$,
and metric $g_{\mu\nu}=\operatorname{diag}(+1,-1,\ldots,-1)$.
By $\R^m$ we denote Euclidean $m$--space
with spatial coordinates $(x^1,\ldots,x^m)$,
and metric $\operatorname{diag}(-1,\ldots,-1)$.
We use Greek letters for tensor indices (irrespectively
of whether we are in Minkowski or Euclidean space),
and repeated indices imply summation.
For vectors in $\M^n$ or $\R^m$ we denote
$\langle u,v\rangle:=u_\mu v^\mu$.
Bold type indicates
the contravariant representation of a Euclidean vector:
${\bold u}=(u^1,\ldots,u^m)=(-u_1,\ldots,-u_m)$.
We write
${\bold u}\boldcdot{\bold v}:=u^\mu v^\mu=-\langle u,v\rangle$.

We use the ``overline'' to denote complex conjugation,
$\,\phantom{\psi}^T\,$ to denote transposition,
and $\,\phantom{\psi}^*\,$ to denote Hermitian conjugation
(combination of complex conjugation and transposition).
Thus, the Dirac
conjugate of a bispinor $\psi$ in our notation is $\psi^*\gamma^0$.
The norm sign without any additional indices indicates the
$L^2$--norm of a function; say, for a vector function
${\bold u}:\R^m\to{\C}^m$ we have
$\|{\bold u}\|^2=\int\overline{\bold u}\boldcdot{\bold u}\,d^mx\,$.

By $m$ and $e=-|e|$ we denote the mass and the charge of the
electron, respectively. The fine structure constant is
$\alpha=e^2\approx\textstyle{\frac1{137}}$.
By $A=(A^0,{\bold A})$ we denote
the electromagnetic vector potential,
which is a given real valued vector function.
Here $A^0\equiv\Phi$ is the electric potential, and
${\bold A}$ is the magnetic vector potential.
We denote $\nabla_\mu:=\partial/\partial x^\mu$ and
$P:=i\nabla-eA$.

Our system of units is such that the speed of light $c$ and
Planck's constant
$\hbar$  are both 1, whereas $m\sim1$
(equivalently, Compton's wavelength $\sim1$).

By $e^{\lambda\mu\nu}$ and $e^{\varkappa\lambda\mu\nu}$
we denote the totally antisymmetric pseudotensor in $\M^3$
and $\M^4$, respectively; we take $e^{012}=e^{0123}=+1$.
In $\M^3$ we define the dual of an antisymmetric tensor $T$ as
${T^\star}_{\lambda}:=
\half e_{\lambda\mu\nu}T^{\mu\nu}$,
and in $\M^4$ as
${T^\star}_{\varkappa\lambda}:=
\half e_{\varkappa\lambda\mu\nu}T^{\mu\nu}$.
In $\M^3$ we define the vector product as
$[v,w]^\lambda:=e^{\lambda\mu\nu}v_\mu w_\nu$;
accordingly,
$\operatorname{curl}:=[\nabla,\ \cdot\ ]$\,, and
$\operatorname{curl}_A:=[\nabla+ieA,\ \cdot\ ]=-i[P,\ \cdot\ ]$\,.
We define divergence on vectors and tensors as
$\operatorname{div}u:=\nabla_\mu u^\mu$ and
$(\operatorname{div}T)^\nu:=\nabla_\mu T^{\mu\nu}$, respectively.

``Pseudo'' refers to quantities which behave in the ``correct'' way
(as scalars, vectors, or tensors) under changes of coordinates
preserving orientation, and which get the ``wrong''
sign under change of orientation.

By $\boldsigma=(\sigma^1,\sigma^2,\sigma^3)$ we denote the
``vector'' of Pauli matrices,
\[
\sigma^1=\pmatrix0&1\\1&0\endpmatrix,\qquad
\sigma^2=\pmatrix0&-i\\ i&0\endpmatrix,\qquad
\sigma^3=\pmatrix1&0\\0&-1\endpmatrix,
\]
and by $\gamma=(\gamma^0,\gamma^1,\gamma^2,\gamma^3)$ the
``vector'' of Dirac matrices.
In Sections 3 and 4 we use the
spinor representation of bispinors as opposed to the
more common standard representation
(see Section 21 of \cite{LL4} for details);
consequently the Dirac matrices appearing in these sections are
\[
\gamma^0=\pmatrix0&I\\ I&0\endpmatrix,\qquad
\gamma^\mu=\pmatrix0&-\sigma^\mu\\
\sigma^\mu&0\endpmatrix,\quad\mu=1,2,3.
\]

\newsection{A tensor interpretation of bispinors}

A bispinor in $\M^4$ is a set of four complex numbers
\begin{equation}\label{bispinor}
\psi=\pmatrix\xi\\\eta\endpmatrix=
\pmatrix\xi^1\\\xi^2\\\eta_{\dot1}\\\eta_{\dot2}
\endpmatrix
\end{equation}
which change under passive Lorentz transformations in the following way.
Under proper Lorentz transformations
$\xi'=B\xi$ and $\eta'=(B^*)^{-1}\eta$,
where $B$ is a complex $2\!\times\!2$ matrix given by
$$
B=\exp\left(-\half\phi\,{\bold n}\boldcdot\boldsigma\right)
=\cosh(\half\phi)-{\bold n}\boldcdot\boldsigma\sinh(\half\phi)
$$
for a boost of speed $V\!=\!\tanh(\phi)$ along
the unit vector ${\bold n}\in\R^3$, and by
\[
B=\exp\left(\half i\vartheta\,{\bold n}\boldcdot\boldsigma\right)
=\cos(\half\vartheta)+
i{\bold n}\boldcdot\boldsigma\sin(\half \vartheta)
\]
for a rotation by angle $\vartheta$ around ${\bold n}$. Here
``passive'' means that we transform the coordinate system
and not the the bispinors themselves, and the
prime refers to the representation of the bispinor
in the new coordinate system.
Under space inversion $\xi'=i\eta$ and $\eta'=i\xi$.
Under time inversion
$\xi'=\sigma^2\overline\xi$,
$\eta'=\sigma^2\overline\eta$.

\begin{remark}\label{remark1}
A full rotation
of the coordinate system ($\vartheta=2\pi$) changes the sign
of a bispinor,
which means that bispinors are defined up to a choice of sign.
\end{remark}

Throughout this section we assume that
we are dealing with a bispinor satisfying the technical condition
$\eta^*\xi\ne0$.

Let $\frak t$ be a fixed real time--like vector and let
$\{e^{(0)},e^{(1)},e^{(2)},e^{(3)}\}$,
$\langle e^{(k)},e^{(l)}\rangle=g^{kl}$,
be some coordinate basis. Set
$\tau:=\operatorname{sign}\langle{\frak t},e^{(0)}\rangle$.
In other words, to every coordinate system we assign a number $\tau=+1$ or
$\tau=-1$. We assume that for the original coordinate system
(and, consequently, for all coordinate systems obtained from the
original one by proper Lorentz transformations and space inversion)
$\tau=+1$. Thus, the vector $\frak t$ fixes the positive direction of time.

Put
\vspace{-5pt}
\begin{equation}\label{rhotheta}
\rho:=|2\eta^*\xi|,\qquad\theta:=\arg(2\eta^*\xi),
\end{equation}
\vspace{-20pt}
\begin{equation}\label{ju}
j^\mu:=\psi^*\gamma^0\gamma^\mu\psi,\qquad
u^\mu:=-i\psi^T\gamma^0\gamma^2\gamma^\mu\psi,
\end{equation}
\vspace{-20pt}
\begin{equation}\label{f}
f^{(0)}:=\frac{\tau j}\rho\,,\qquad
f^{(1)}:=\frac{\operatorname{Re}u}\rho\,,\qquad
f^{(2)}:=\frac{\tau\operatorname{Im}u}\rho\,.
\end{equation}

\begin{theorem}\label{theorem1}
The quantities
\begin{equation}\label{tensor}
\{\rho,\,\theta;\,f^{(k)},\ k=0,1,2\}
\end{equation}
defined in accordance with (\ref{rhotheta})--(\ref{f})
have the following properties:

$\rho\in\R_+$ is a scalar;

$\theta\in{\Bbb S}^1$ is a pseudoscalar;

the $f^{(k)}$ are real vectors
forming an orthonormal triad, i.e.,
$\langle f^{(k)},f^{(l)}\rangle=g^{kl}$.

Conversely, given a set (\ref{tensor}) with the above
three properties
there is a unique bispinor (\ref{bispinor}) satisfying
(\ref{rhotheta})--(\ref{f}).
\end{theorem}

Of course, ``uniqueness'' of a bispinor is understood as
``uniqueness up to the choice of sign''; see Remark \ref{remark1}.

\begin{proofo} of Theorem \ref{theorem1}.
Clearly $f^{(1)}$ and $f^{(2)}$ are real whilst
$\big(\gamma^0\gamma^\mu\big)^*={\gamma^\mu}^*\gamma^0
=\gamma^0\gamma^\mu$ from which it follows that $j$ and hence
$f^{(0)}$ are also real.

Given a Lorentz transformation $\Lambda$ of the coordinate system let
$\Lambda^\mu{}_\nu$, $\mu,\nu=0,1,2,3$,
be the real $4\!\times\!4$ matrix in
terms of which the vector $a$ transforms to the
vector $a'$ with components given by
\[
{a'}^\mu=\Lambda^\mu{}_\nu a^\nu.
\]
For a boost of speed $V\!=\!\tanh(\phi)$ along
the unit vector ${\bold n}\in\R^3$
\[
\Lambda=\exp\big(-\phi M_{\bold n}\big)
=I+\big(\cosh(\phi)-1\big)M_{\bold n}^2-\sinh(\phi)M_{\bold n},
\]
where
\[
M_{\bold n}=
\pmatrix
0&n^1&n^2&n^3\\ n^1&0&0&0\\ n^2&0&0&0\\ n^3&0&0&0
\endpmatrix,
\]
whilst, for a rotation by angle $\vartheta$ around $\mathbf n$,
\[
\Lambda=\exp\big(-\vartheta N_{\bold n}\big)
=I+\big(1-\cos(\vartheta)\big)N_{\bold n}^2
-\sin(\vartheta)N_{\bold n},
\]
where
\[
N_{\bold n}=
\pmatrix
0&0&0&0\\ 0&0&-n^3&n^2\\ 0&n^3&0&-n^1\\ 0&-n^2&n^1&0
\endpmatrix.
\]

If the Lorentz transformation $\Lambda$ is proper the corresponding
transformation of the bispinor $\psi$ is linear and so
can be represented by a complex $4\!\times\!4$ matrix $S(\Lambda)$;
that is $\psi'=S(\Lambda)\psi$. Clearly
\[
S(\Lambda)=
\pmatrix B(\Lambda)&0\\ 0&\big(B^*(\Lambda)\big)^{-1}\endpmatrix,
\]
where $B(\Lambda)$ is the complex $2\!\times\!2$ matrix defined above.
A straightforward check using the explicit
forms of the matrices $S(\Lambda)$ and $B(\Lambda)$ for boosts and spatial
rotations gives us
\begin{equation}
\label{relSgamma}
\Lambda^\mu{}_\nu\gamma^\nu=S^{-1}(\Lambda)\gamma^\mu S(\Lambda)
\end{equation}
in these cases (see Section 7.2 of \cite{S} for more details).
Formula (\ref{relSgamma}) immediately extends to all proper Lorentz
transformations since any such transformation can be written as a
product of boosts and rotations.
On the other hand $\det\big(B(\Lambda)\big)=1$ from which we get
$\sigma^2B^{-1}(\Lambda)=B^T(\Lambda)\sigma^2$ and
$\sigma^2B^*(\Lambda)=\overline{B(\Lambda)}^{\;-1}\sigma^2$.
Hence
\begin{equation}\label{relSgamma0}
S^*(\Lambda)\gamma^0=\gamma^0S^{-1}(\Lambda),
\end{equation}
\vspace{-20pt}
\begin{equation}\label{relSgamma2}
\overline{S(\Lambda)}^{\;-1}\gamma^2=\gamma^2S^{-1}(\Lambda).
\end{equation}

Now, for proper Lorentz transformations,
\[
{\eta'}^*\xi'
=\big(\big(B^*(\Lambda)\big)^{-1}\eta\big)^*B(\Lambda)\xi
=\eta^* B^{-1}(\Lambda)B(\Lambda)\xi=\eta^*\xi,
\]
so $\rho'=\rho$ and $\theta'=\theta$. Also, by
(\ref{relSgamma}) and (\ref{relSgamma0}),
\begin{eqnarray*}
\lefteqn{{j'}^\mu={\psi'}^*\gamma^0\gamma^\mu\psi'
=\psi^* S^*(\Lambda)\gamma^0\gamma^\mu S(\Lambda)\psi}\quad\\
&&{}=\psi^* \gamma^0 S^{-1}(\Lambda)\gamma^\mu S(\Lambda)\psi
=\psi^* \gamma^0 \Lambda^\mu{}_\nu\gamma^\nu \psi
=\Lambda^\mu{}_\nu j^\nu,
\end{eqnarray*}
whilst a similar argument using (\ref{relSgamma2}) as well
gives
\begin{eqnarray*}
\lefteqn{{u'}^\mu
=-i\psi^T S^T(\Lambda)\gamma^0\gamma^2\gamma^\mu S(\Lambda)\psi}
\quad\\
&&{}=-i\psi^T \gamma^0 \overline{S(\Lambda)}^{\;-1}\gamma^2\gamma^\mu
S(\Lambda)\psi
=-i\psi^T\gamma^0\gamma^2 S^{-1}(\Lambda)\gamma^\mu S(\Lambda)\psi
=\Lambda^\mu{}_\nu u^{\nu}.
\end{eqnarray*}

Under space inversion $\xi'=i\eta$ and $\eta'=i\xi$ so
${\eta'}^*\xi'=\xi^*\eta=\overline{\eta^*\xi}$
and $\psi'=S(\Lambda)\psi$ where
$S(\Lambda)=\pmatrix 0&iI\\ iI&0 \endpmatrix$.
Thus $\rho'=\rho$ and $\theta'=-\theta$. On the other hand
$\Lambda^\mu{}_\nu=\operatorname{diag}(+1,-1,-1,-1)$
which can be used to directly check that (\ref{relSgamma})
to (\ref{relSgamma2}) are still valid; it follows that
${j'}^\mu=\Lambda^\mu{}_\nu j^\nu$ and
${u'}^\mu=\Lambda^\mu{}_\nu u^\nu$.

Under time inversion $\xi'=\sigma^2\overline{\xi}$
and $\eta'=\sigma^2\overline{\eta}$ so
${\eta'}^*\xi'=\overline{\eta}^*\big(\sigma^2\big)^*
\sigma^2\overline{\xi}=\overline{\eta^*\xi}$, giving $\rho'=\rho$ and
$\theta'=-\theta$ once again.
Now $\Lambda^\mu{}_\nu=\operatorname{diag}(-1,+1,+1,+1)$.
We cannot extend the definition of
$S(\Lambda)$ to cover the present case; however if we set
$S_t=\pmatrix \sigma^2&0 \\ 0&\sigma^2 \endpmatrix$ then
\[
\psi'=S_t\overline{\psi},\quad \Lambda^\mu{}_\nu\overline{\gamma^\nu}
=-S_t^{-1}\gamma^\mu S_t,\quad S_t^*\gamma^0=\gamma^0 S_t^{-1}
\quad\mbox{and}\quad \overline{S_t}^{\;-1}\gamma^2
=\overline{\gamma^2}S_t^{-1}.
\]
Therefore
\[
{j'}^\mu=\psi^T S_t^*\gamma^0\gamma^\mu
S_t\overline{\psi}
=-\psi^T\gamma^0\Lambda^\mu{}_\nu\overline{\gamma^\nu}\,\overline{\psi}
=-\Lambda^\mu{}_\nu\overline{j^\nu}=-\Lambda^\mu{}_\nu j^\nu,
\]
(n.b.~$\gamma^0$ and $j$ are real) and
\begin{eqnarray*}
\lefteqn{{u'}^\mu
=-i\psi^* S_t^T\gamma^0\gamma^2\gamma^\mu S_t\overline{\psi}}
\quad\\
&&{}=-i\psi^* \gamma^0 \overline{S_t}^{\;-1}\gamma^2\gamma^\mu
S_t\overline{\psi}
=i\psi^*\gamma^0\overline{\gamma^2}
\Lambda^\mu{}_\nu\overline{\gamma^\nu}\,\overline{\psi}
=\Lambda^\mu{}_\nu\overline{u^\nu}.
\end{eqnarray*}
However $\tau'=-\tau$ so
${{f^{(k)}}'}^\mu=\Lambda^\mu{}_\nu {f^{(k)}}^\nu$ for $k=0,1,2$.

Since any Lorentz transformation can be written
as a combination of a proper transformation,
space inversion and time inversion,
the above calculations show that $\rho$ is a scalar,
$\theta$ is a pseudoscalar and $f^{(k)}$ is a vector for $k=0,1,2$.

The fact that $\{f^{(k)},\ k=0,1,2\}$ is an orthonormal triad can be
checked directly by a somewhat lengthy calculation. However, since we
have established that the $f^{(k)}$'s are vectors, it suffices to
prove they are orthonormal for one particular choice of
the coordinate system.

\emph{Claim: there exists a
unique proper coordinate system in which}
\begin{equation}\label{specialpsi}
\psi=\pmatrix \xi\\ \eta\endpmatrix
=\pm\sqrt{\frac\rho2}
\pmatrix e^{i\theta/2}\\ 0\\ e^{-i\theta/2}\\ 0\endpmatrix
\end{equation}
\emph{for some $\rho\in\R_+$ and $\theta\in{\Bbb S}^1$.
Of course, these $\rho$ and $\theta$ must be given
by (\ref{rhotheta}).}

Indeed, we have
\vspace{-5pt}
\begin{eqnarray*}
\lefteqn{j^\mu j_\mu
=\big(\xi^*\xi+\eta^*\eta\big)^2
-\sum_{\nu=1}^3\big(\xi^*\sigma^\nu\xi-\eta^*\sigma^\nu\eta\big)^2}
\quad\\
&&=4\big(\overline{\xi^1}\eta_{\dot1}+\overline{\xi^2}\eta_{\dot2}\big)
\big(\xi^1\overline{\eta_{\dot1}}+\xi^2\overline{\eta_{\dot2}}\big)
=\big|2\eta^*\xi\big|^2=\rho^2>0.
\end{eqnarray*}
Therefore the real vector $j$ is time-like so we can choose a
proper coordinate system, given by an appropriate boost,
in which $\bold j=0$ and $j^0=\rho$;
\,n.b.~$j^0=\xi^*\xi+\eta^*\eta>0$.
Now suppose we rotate the coordinate system by an
angle $\vartheta\in\R$ about the unit vector
${\bold n}=(n^1,n^2,0)\in\R^3$.
From above we have $\xi'=B\xi$ where
\[
B=\pmatrix \cos(\half\vartheta)& (n^2+in^1)\sin(\half \vartheta)\\
\Big.(-n^2+in^1)\sin(\half \vartheta)& \cos(\half\vartheta)
\endpmatrix.
\]
Choosing $n^1$, $n^2$ and $\vartheta$ so that
$\,(-n^2+in^1)\sin(\half \vartheta)\xi^1
+\cos(\half\vartheta)\xi^2=0\;$
we thus have ${\xi'}^2=0$. Since we are only rotating the coordinate
system we must have $\bold j'=0$ and ${j'}^0=\rho$. An elementary
analysis of these formulae gives us
${\eta'}_{\dot2}=0$ and
$|{\eta'}_{\dot1}|=\big|{\xi'}^1\big|=\sqrt{\frac\rho2}$.

Suppose now that we are in the rotated coordinate system (and drop the
primes). Rotate this coordinate system by an angle $\vartheta\in\R$ about
${\bold n}=(0,0,1)$. From above we have $\xi'=B\xi$ and
$\eta'=(B^*)^{-1}\eta$ where
\[
B=\pmatrix\alpha&0\\0&\overline{\alpha}\endpmatrix,
\qquad\alpha=e^{i\vartheta/2}.
\]
Therefore
$\,{\xi'}^2={\eta'}_{\dot2}=0$,
$\,{\xi'}^1=\alpha\xi^1$ and
$\,{\eta'}_{\dot1}=\alpha\eta_{\dot1}$. Now,
$|\overline{\eta_{\dot1}}|=|{\xi^1}|$
so we can choose $\vartheta$ such that
${\xi'}^1=\overline{{\eta'}_{\dot1}}$. However
${\eta'}^*\xi'=\eta^*\xi$ (since we have only changed the coordinate
system by proper Lorentz transformations) so
$2{\eta'}^*\xi'=2\bigl({\xi'}^1\bigr)^2
=2\bigl(\,\overline{{\eta'}_{\dot1}}\,\bigr)^2=\rho e^{i\theta}$,
giving us the required representation (\ref{specialpsi}).
In this coordinate system it is straightforward
to check that
\begin{equation}\label{specialf}
f^{(0)}=\pmatrix 1\\ 0\\ 0\\ 0\endpmatrix,\qquad
f^{(1)}=\pmatrix 0\\ 1\\ 0\\ 0\endpmatrix,\qquad
f^{(2)}=\pmatrix 0\\ 0\\ 1\\ 0\endpmatrix
\end{equation}
(contravariant representation). Clearly, (\ref{specialf})
can be true only in one proper coordinate system.
This completes the proof of the claim.

Formula (\ref{specialf}) shows that $\{f^{(k)},\ k=0,1,2\}$
is an orthonormal triad. The proof of the direct statement
of Theorem \ref{theorem1} is complete.

Conversely, suppose we have a set
(\ref{tensor}) with the required properties.
Choose the proper coordinate system in
which (\ref{specialf}) holds and define the bispinor $\psi$
by (\ref{specialpsi}). It follows that this $\psi$ satisfies
(\ref{rhotheta})--(\ref{f}). Now, let $\widetilde\psi$
be another bispinor satisfying (\ref{rhotheta})--(\ref{f}).
The above claim associates with $\widetilde\psi$ a special
proper coordinate system in which $\widetilde\psi$ is
given by the right--hand side of (\ref{specialpsi}). But
(\ref{specialf}) can be true only in one proper coordinate system,
 so the special coordinate system for the bispinor
$\widetilde\psi$ is the one we are already working in.
This implies $\widetilde\psi=\psi$.
\end{proofo}

Set ${f^{(3)}}^\nu\!:=e^{\varkappa\lambda\mu\nu}
{f^{(0)}}_\varkappa{f^{(1)}}_\lambda{f^{(2)}}_\mu$\,.
We shall call the pseudovector
${f^{(3)}}$ {\em spin.\/} Clearly,
$\{f^{(k)},\ k=0,1,2,3\}$ is an orthonormal tetrad
(orthonormal frame).

\newsection{The Dirac equation in tensor form}

From now on our bispinor $\psi$ is a function of the point
$x$ in space--time. We assume that $\psi$ is defined on some open set
$O\subset\M^4$, and that $\psi$ is smooth and satisfies the technical
condition $\eta^*\xi\ne0$. Our objective is to rewrite
the Dirac equation (\ref{Dirac})
in terms of the tensor quantities (\ref{tensor}).

It will be convenient for us to deal with
Euler--Lagrange functionals rather than with
the corresponding differential equations. It is well known that
the Euler--Lagrange functional for (\ref{Dirac}) is
\begin{equation}\label{EulerDirac}
\int\psi^*\gamma^0\bigl(
\gamma^\varkappa(i\nabla\!-\!\tau eA)_\varkappa-m
\bigr)\psi\,d^4x
\end{equation}
(we included the factor $\tau$ because the Dirac equation is
$CT$--invariant, but not $T$--invariant).
Integration in (\ref{EulerDirac}) is carried out over $O$.
We do not assume, however, that the integral converges
and treat (\ref{EulerDirac}) as a formal expression.
This is acceptable for our purposes because we are interested
not in (\ref{EulerDirac}) itself, but only
in its variation generated by a
variation of the bispinor $\delta\psi\in C_0^\infty(O)$.
The same applies to all subsequent functionals.

Further on we use
\begin{equation}\label{EulerDirac1}
\int\operatorname{Re}\Bigl(\psi^*\gamma^0\bigl(
\gamma^\varkappa(i\nabla\!-\!\tau eA)_\varkappa-m
\bigr)\psi\Bigr)d^4x
\end{equation}
instead of (\ref{EulerDirac}) as the Euler--Lagrange functional
for (\ref{Dirac}). This is possible because the variation of
(\ref{EulerDirac}) is real.
Incidentally, without the $\,\operatorname{Re}\,$ the integrand
is not a scalar (time inversion leads to complex conjugation).

Let us now examine the rotation of the orthonormal frame
$\{f^{(k)},\ k=0,1,2,3\}$ as we move from one point $x$ to another.
Let
\begin{equation}\label{rotation1}
\delta{f^{(k)}}_\mu=\bigl(\nabla_\nu {f^{(k)}}_\mu\bigr)\delta x^\nu
\end{equation}
be the increment of the vector function $f^{(k)}$ when we
move from $x$ to a close point $x+\delta x$.
We define the (antisymmetric) tensor of infinitesimal rotations
$R$ as the solution of the linear system
\begin{equation}\label{rotation2}
{\delta f^{(k)}}_\mu=R_{\mu\lambda}{f^{(k)}}^\lambda,
\qquad k=0,1,2,3.
\end{equation}
The explicit formula for the solution of (\ref{rotation2}) is
\begin{equation}\label{rotation3}
R_{\mu\lambda}=\sum_{j,l=0}^3
g_{jl}\bigl({\delta f^{(j)}}_\mu\bigr){f^{(l)}}_\lambda\,.
\end{equation}
In particular, formulae (\ref{rotation3}), (\ref{rotation1})
imply
\begin{equation}\label{rotation4}
\bigl(\operatorname{div}(R^\star)\bigr)^\varkappa=
\frac12\sum_{j,l=0}^3 g_{jl}e^{\epsilon\varkappa\mu\lambda}
\bigl({\nabla_\epsilon f^{(j)}}_\mu\bigr){f^{(l)}}_\lambda\,.
\end{equation}
As $R^\star$ is a pseudotensor,
$\operatorname{div}(R^\star)$ is a pseudovector.

\begin{theorem}\label{theorem2}
We have the identity
\begin{multline}\label{identity}
\operatorname{Re}\Bigl(\psi^*\gamma^0\bigl(
\gamma^\varkappa(i\nabla\!-\!\tau eA)_\varkappa-m
\bigr)\psi\Bigr)\,=\\
-
\left[
\frac12
\bigl\langle f^{(3)},\,\operatorname{div}(R^\star)+
\operatorname{grad}\theta\bigr\rangle
\,+\,e\bigl\langle f^{(0)},A\bigr\rangle\,+\,m\cos\theta
\right]
\!\rho\,.
\end{multline}
\end{theorem}

\begin{proofo} of Theorem \ref{theorem2}.
Since both sides of (\ref{identity}) are scalars it suffices
to check the identity at each point $x\in O$ in only one coordinate
system. Choosing the coordinate system given by
the claim in the proof of
Theorem \ref{theorem1} (based at the point $x$) we have
\[
f^{(k)}{}_\mu=g_{k\mu},\qquad k,\mu=0,1,2,3;
\]
n.b.~this, and other expressions to follow, are \emph{not} tensor
identities but rather identities which hold at the point $x$ for our
special choice of coordinates. Using this expression and
(\ref{rotation4}) we have
\[
\bigl(\operatorname{div}(R^\star)\bigr)_3
=\nabla_0f^{(1)}{}_2-\nabla_1f^{(0)}{}_2+\nabla_2f^{(0)}{}_1\,,
\]
and hence
\begin{eqnarray*}
\lefteqn{\rho\bigl\langle f^{(3)},\,\operatorname{div}(R^\star)
+\operatorname{grad}\theta\bigr\rangle
=\rho\bigl(\nabla_0f^{(1)}{}_2-\nabla_1f^{(0)}{}_2
+\nabla_2f^{(0)}{}_1+\nabla_3\theta\bigr)}\\
&&{}=\nabla_0\big(\rho f^{(1)}{}_2\big)
-\nabla_1\big(\rho f^{(0)}{}_2\big)
+\nabla_2\big(\rho f^{(0)}{}_1\big)
-ie^{-i\theta}\nabla_3(\rho e^{i\theta})+i\nabla_3\rho\,.
\end{eqnarray*}
Combining the general definitions
(\ref{rhotheta})--(\ref{f})
with the specific value of the bispinor $\psi(x)$ in our chosen
coordinate frame (formula (\ref{specialpsi})) we get
\begin{eqnarray*}
\nabla_0\bigl(\rho f^{(1)}{}_2\bigr)
&=&i\Bigl({}-\overline{\xi^1}\nabla_0\xi^1
+\xi^1\nabla_0\overline{\xi^1}
-\overline{\eta_{\dot1}}\nabla_0\eta_{\dot1}
+\eta_{\dot1}\nabla_0\overline{\eta_{\dot1}}\Bigr)\,,\\
\nabla_1\bigl(\rho f^{(0)}{}_2\bigr)
&=&i\Bigl(\overline{\xi^1}\nabla_1\xi^2
-\xi^1\nabla_1\overline{\xi^2}
-\overline{\eta_{\dot1}}\nabla_1\eta_{\dot2}
+\eta_{\dot1}\nabla_1\overline{\eta_{\dot2}}\Bigr)\,,\\
\nabla_2\bigl(\rho f^{(0)}{}_1\bigr)
&=&{}-\overline{\xi^1}\nabla_2\xi^2
-\xi^1\nabla_2\overline{\xi^2}
+\overline{\eta_{\dot1}}\nabla_2\eta_{\dot2}
+\eta_{\dot1}\nabla_2\overline{\eta_{\dot2}}\,,\\
e^{-i\theta}\nabla_3\bigl(\rho e^{i\theta}\bigr)
&=&2\overline{\xi^1}\nabla_3\xi^1
+2\eta_{\dot1}\nabla_3\overline{\eta_{\dot1}}\,,\\
\nabla_3\rho
&=&\overline{\xi^1}\nabla_3\xi^1
+\xi^1\nabla_3\overline{\xi^1}
+\overline{\eta_{\dot1}}\nabla_3\eta_{\dot1}
+\eta_{\dot1}\nabla_3\overline{\eta_{\dot1}}\,.
\end{eqnarray*}
The above expressions can then be combined producing
\begin{eqnarray*}
\lefteqn{\rho\bigl\langle f^{(3)},\,\operatorname{div}(R^\star)
+\operatorname{grad}\theta\bigr\rangle}\quad\\
&=&-2\operatorname{Re}\Big\{\overline{\xi^1}
\bigl(i\nabla_0\xi^1+i\nabla_1\xi^2+\nabla_2\xi^2+i\nabla_3\xi^1
\bigr)\\
&&\qquad
{}+\overline{\eta_{\dot1}}
\bigl(i\nabla_0\eta_{\dot1}-i\nabla_1\eta_{\dot2}
-\nabla_2\eta_{\dot2}-i\nabla_3\eta_{\dot1}\bigr)\!\Big\}\\
&=&-2\operatorname{Re}\left\{
\pmatrix\xi^*& \!\!\!\eta^*\endpmatrix
\left[\pmatrix I&0\\ 0&I\endpmatrix i\nabla_0
+\sum_{\nu=1}^3
\pmatrix \sigma^\nu&0\\ 0&\!\!\!-\sigma^\nu\endpmatrix i\nabla_\nu
\right]\pmatrix \xi\\ \eta\endpmatrix\right\}\\
&=&-2\operatorname{Re}
\left(\psi^*\gamma^0\gamma^\mu i\nabla_\mu\psi\right).
\end{eqnarray*}

From the definition of $f^{(0)}$ we have
\[
\rho e\langle f^{(0)},A\rangle
=\rho {f^{(0)}}^\mu eA_\mu
=\psi^*\gamma^0\gamma^\mu\tau eA_\mu\psi,
\]
whilst the definition of $\rho$ and $\theta$ gives
\[
\rho m\cos\theta=m\operatorname{Re}(2\eta^*\xi)
=m\pmatrix\xi^*& \!\!\!\eta^*\endpmatrix
\pmatrix 0&I\\ I&0\endpmatrix
\pmatrix \xi\\ \eta\endpmatrix
=m\psi^*\gamma^0\psi.
\]
The result now follows from the fact that both of these expressions
are real.
\end{proofo}

Formulae (\ref{EulerDirac1}) and (\ref{identity}) imply
\begin{corollary}\label{corollary1}
The Euler--Lagrange functional for the Dirac equation can be
written as
\begin{equation}\label{EulerDirac2}
\int\left[
\frac12
\bigl\langle f^{(3)},\,\operatorname{div}(R^\star)+
\operatorname{grad}\theta\bigr\rangle
\,+\,e\bigl\langle f^{(0)},A\bigr\rangle\,+\,m\cos\theta
\right]
\!\rho\,d^4x\,.
\end{equation}
\end{corollary}

Variation of (\ref{EulerDirac2}) with respect to
the scalar $\rho$, pseudoscalar $\theta$ and
the moving frame $\{f^{(k)},\ k=0,1,2,3\}$
produces a (nonlinear) system of tensor differential equations
equivalent to the Dirac equation (\ref{Dirac}). Of course, in
performing this variation one has to remember the constraints:
all the quantities are real, $\rho$ is positive, and
$\langle f^{(k)},f^{(l)}\rangle=g^{kl}$.

\newsection{Simplifications in the case of $\M^3$}

Let us return to the Dirac equation (\ref{Dirac}) and consider the
case when $A$ and $\psi$ do not depend on $x^3$. Then
(\ref{Dirac}) separates into two systems
of two equations :
\begin{equation}
\label{DiracM3}
\pmatrix
P_0&P_\mp\\
-P_\pm&-P_0
\endpmatrix
\pmatrix\varphi_\pm\\\chi_\pm\endpmatrix=
m\pmatrix\varphi_\pm\\\chi_\pm\endpmatrix,
\end{equation}
where $P_\pm:=P_1\pm iP_2$ and
\[
\pmatrix\varphi_+\\\varphi_-\endpmatrix=\frac{\xi+\eta}{\sqrt2}\,,
\qquad
\pmatrix\chi_-\\\chi_+\endpmatrix=\frac{\xi-\eta}{\sqrt2}\,.
\]
(The relation with standard notation is
$\ \varphi_+=\varphi_1$, $\ \varphi_-=\varphi_2$,
$\ \chi_-=\chi_1$, $\ \chi_+=\chi_2$;
see formula (21.17) in \cite{LL4}.)
Accordingly, simplifications occur in the tensor functional
(\ref{EulerDirac2}). We give the final result omitting
intermediate calculations.

We are now working in $\M^3$ and the particle
is described by the set of quantities
\begin{equation}
\label{tensorM3}
\{\rho;\,f^{(k)},\ k=1,2\}
\end{equation}
where $\rho\in\R_+$ is a scalar and
the $f^{(k)}$ are real vectors
forming an orthonormal dyad, i.e.,
$\langle f^{(k)},f^{(l)}\rangle=g^{kl}$.
As we are in a 3--space it is convenient to use the notion of
a vector product. Put $f^{(0)}:=[f^{(1)},f^{(2)}]$,
and define the pseudovector of infinitesimal rotations
$r$ as the solution of the linear system
$\delta f^{(k)}=[r,f^{(k)}]$, $\,k=0,1,2$.
Then $r=-R^\star$ where
\begin{equation}
\label{rotation3M3}
R_{\mu\lambda}=\sum_{j,l=0}^2
g_{jl}\bigl({\delta f^{(j)}}_\mu\bigr){f^{(l)}}_\lambda
\end{equation}
(cf.~(\ref{rotation3})), and
\begin{equation}
\label{rotation4M3}
\operatorname{div}r=-\frac12\sum_{j,l=0}^2
g_{jl}\langle f^{(j)},\operatorname{curl}f^{(l)}\rangle
\end{equation}
(cf.~(\ref{rotation4})).
The functional \eqref{EulerDirac2} turns into
\begin{equation}
\label{EulerDirac2M3}
\int\left[
\frac12\operatorname{div}r
+e\bigl\langle f^{(0)},A\bigr\rangle\pm m
\right]\!\rho\,d^3x\,,
\end{equation}
where the sign corresponds to that in (\ref{DiracM3}).
The functional(s) (\ref{EulerDirac2M3}) should be varied with respect
to $\rho$ and the moving frame $\{f^{(k)},\ k=0,1,2\}$.

\begin{remark}\label{remark2}
The $f^{(0)}$ from this section is a pseudovector in $\M^3$, and
it coincides up to sign with the corresponding part of the 4--vector
$f^{(0)}$ from Sections 3 and 4.
\end{remark}

\begin{remark}\label{remark3}
The set (\ref{tensorM3}) is equivalent to
a complex valued vector function $u$
satisfying the constraint $\langle u,u\rangle=0$;
the equivalence is established by the formula
$u=\rho(f^{(1)}+if^{(2)})$, cf.~(\ref{f}).
This is not surprising: Cartan originally
defined spinors as
complex vectors $u$ satisfying $\langle u,u\rangle=0$,
see Section 52 in \cite{C}.
\end{remark}

\newsection{A model equation in $\M^3$}

The arguments in this section are not mathematically rigorous,
and are needed to motivate the introduction of the equation
(\ref{model}).

Let $\alpha\to+0$ be an asymptotic parameter. Assume
that the external electromagnetic field $A$ is smooth
and satisfies
\begin{equation}
\label{scaling}
eA\sim\alpha^2,\qquad
\partial_x^\beta A\sim\alpha^{|\beta|}A\,,
\end{equation}
where $\beta=(\beta_0,\ldots,\beta_3)$ is an
arbitrary multiindex,
$|\beta|=\beta_0+\ldots+\beta_3$,
$\,\partial_x^\beta A=\partial_x^\beta A^\mu=
(\nabla_0)^{\beta_0}\ldots(\nabla_3)^{\beta_3}A^\mu$,
and ``$\sim$'' stands for ``asymptotically of the order of''.
In other words, we assume that the field is weak
(potential energy of the electron $\sim\alpha^2$)
and slowly varying
(each differentiation gives an additional $\alpha$).

Our scaling assumptions (\ref{scaling}) are meant to
model the situation which occurs in the hydrogen
or positronium atoms.
Indeed for the hydrogen atom
\begin{equation}
\label{hydrogen1}
e A^0=e\Phi=-\alpha r^{-1},\qquad{\bold A}\equiv{\bold 0},
\end{equation}
where $r=\sqrt{(x^1)^2+(x^2)^2+(x^3)^2}$.
But the characteristic length associated with the
wave functions of bound states is
\begin{equation}
\label{hydrogen2}
r\sim\alpha^{-1}.
\end{equation}
The latter is established by elementary analysis of the
corresponding Schr\"odinger equation; say, the wave function
of the ground state is $\Psi=e^{-\alpha mr}$.
Formulae (\ref{hydrogen1}), (\ref{hydrogen2}) imply
(\ref{scaling}).

Of course, such arguments should
be treated with a fair degree of caution as the
Coulomb potential has a singularity at the origin.
However, in theoretical physics it is common to disregard
this technical difficulty, and it is known
(see, e.g., Sections 33 and 34 in \cite{LL4})
that one can get very sharp results on the basis of
formal asymptotic calculations of the type
(\ref{scaling})--(\ref{hydrogen2}).

The question we address now is whether it is possible to
suggest simple tensor equations which would be
asymptotically equivalent
(up to a certain accuracy in powers of the small parameter
$\alpha$) to the Dirac equation. In our search we accept
equations whose algebraic structure may be totally different
from that of the Dirac equation, as long as they
have (asymptotically) the required spectral properties.

Examination of (\ref{EulerDirac2M3}),
(\ref{rotation4M3}) and Remark \ref{remark3}
suggests (\ref{model}) as the natural candidate in $\M^3$.
Let us rewrite (\ref{model}) as
\begin{equation}
\label{model1}
-i[P,u]=\pm m u\,,
\end{equation}
and formally analyse the properties of this equation. Rigorous
mathematical ana\-ly\-sis is deferred till the next section.

Let us first make some general observations.

\emph{Observation 1: the equation (\ref{model1}) is not
algebraically
equivalent to the Dirac equation (\ref{DiracM3}).} This is clear
from the fact that the number of equations in (\ref{model1})
(3 equations) and (\ref{DiracM3}) (2 equations) is different.
Also, there are no spinors in (\ref{model1}).

\emph{Observation 2: the equation (\ref{model1}) is Lorentz
invariant.} In fact, (\ref{model1}) is probably ``more invariant''
than the Dirac equation because
it can be used in curved space--time:
the notions of vector product and $\operatorname{curl}$ are
defined on any pseudo--Riemannian 3--manifold.

\emph{Observation 3: the equation (\ref{model1}) is formally
self--adjoint:}
\[
\int\big\langle\overline v\,,-i[P,u]\mp mu\big\rangle\,d^3x
\,=\,
\int\big\langle\,\overline{-i[P,v]\mp mv}\,,u\big\rangle\,d^3x\,.
\]
Moreover, it has an Euler--Lagrange functional
which can be written as
\[
\int\big\langle\overline u\,,-i[P,u]\mp mu\big\rangle\,d^3x
\qquad\text{or}\qquad
\int\operatorname{Re}
\big\langle\overline u\,,-i[P,u]\mp mu\big\rangle\,d^3x\,.
\]

Equation (\ref{model1}) can be presented in matrix form as
\[
\pmatrix
0   & P_2 & -P_1\\
P_2 & 0   & P_0\\
-P_1& -P_0& 0
\endpmatrix\pmatrix
u^0\\ u^1\\ u^2
\endpmatrix=\pm im\pmatrix
u^0\\ u^1\\ u^2
\endpmatrix.
\]
The first row gives
$u^0=\mp i\bigl(P_2u^1-P_1u^2\bigr)/m\;$ which can
then be used to eliminate $u^0$ from the remaining two rows.
This results in a $2\!\times\!2$ second order system of
equations which is equivalent to equation (\ref{model1}):
\begin{equation}
\label{model2}
\pmatrix
m^2+P_2{}^2   &\pm imP_0-P_2P_1\\
\mp imP_0-P_1P_2& m^2+P_1{}^2
\endpmatrix\pmatrix
u^1\\ u^2
\endpmatrix=0\,.
\end{equation}

Now, suppose we are looking
for bound state solutions; that is,
assume that $A$ does not depend on $x^0$ and
$u$ is of the form
$u(x^1,x^2)e^{-i\varepsilon x^0}$.
For such vector functions we have $\,P_0=\varepsilon-e\Phi\,$,
and (\ref{model2}) reduces to
\begin{equation}
\label{spectral1}
{\cal A}{\bold u}=\varepsilon{\cal B}{\bold u}
\end{equation}
where
\[
{\cal A}:=\pmatrix m^2+P_2{}^2&\mp ime\Phi-P_2P_1\\
\pm ime\Phi-P_1P_2& m^2+P_1{}^2\endpmatrix,\qquad
{\cal B}:=\pmatrix 0&\mp im\\\pm im& 0\endpmatrix,
\]
\[
{\bold u}=\pmatrix u^1\\ u^2\endpmatrix:
{\Bbb R}^2\longrightarrow{\Bbb C}^2,
\]
and $\varepsilon$ is the spectral parameter.
Note that ${\cal A}$ is not elliptic.

\emph{Observation 4: equation (\ref{spectral1}) asymptotically
reduces to the Pauli equation}.
Indeed, the unitary transformation
\[
\pmatrix u^1\\ u^2\endpmatrix=
\frac1{\sqrt2}
\pmatrix1& 1\\ i&-i\endpmatrix
\pmatrix\varphi_+\\\varphi_-\endpmatrix
\]
turns (\ref{spectral1}) into
\begin{equation}
\label{unitary}
\!\!\!\!\!\!\!\!\!\!\!\!\!\!
\pmatrix
m^2+\frac{P_-P_+}2\pm me\Phi\mp\varepsilon m&
-\frac{{P_-}^2}2\\
-\frac{{P_+}^2}2&
m^2+\frac{P_+P_-}2\mp me\Phi\pm\varepsilon m
\endpmatrix
\pmatrix\varphi_+\\\varphi_-\endpmatrix=0\,.
\end{equation}
The latter system formally reduces to the scalar equation
\[
\left(\frac{P_\mp P_\pm}{2m}+e\Phi
-\frac1{4m^2}{P_\mp}^2
\left(m+\varepsilon+\frac{P_\pm P_\mp}{2m}-e\Phi\right)^{-1}
{P_\pm}^2\right)\varphi_\pm\,
=\,(\varepsilon-m)\varphi_\pm\,.
\]
Suppose we are looking for the bound states of the electron,
so that $\varepsilon\approx+m$,
and suppose that
our eigenfunction inherits the slow variation
property of the potential,
${P_1}^{\beta_1}{P_2}^{\beta_2}\varphi_\pm
\sim\alpha^{\beta_1+\beta_2}\varphi_\pm$.
Then the above scalar equation can be rewritten as
\begin{equation}
\label{Pauli1}
\left(\frac{P_\mp P_\pm}{2m}+e\Phi+O(\alpha^4)
\right)\varphi_\pm\,
=\,(\varepsilon-m)\varphi_\pm\,.
\end{equation}
But $\,P_\mp P_\pm={\bold P}^2\mp eH^3$, where
$\,{\bold P}^2:={P_1}^2+{P_2}^2$, and
$\,H^3=ie^{-1}(P_2P_1-P_1P_2)=\nabla_2 A_1-\nabla_1 A_2\,$
is the intensity of the magnetic field,
see formula (23.5) and list of notation in \cite{LL2}.
Therefore, (\ref{Pauli1}) takes the form
\begin{equation}
\label{Pauli2}
\left(\frac{{\bold P}^2\mp eH^3}{2m}+e\Phi+O(\alpha^4)
\right)\varphi_\pm\,
=\,(\varepsilon-m)\varphi_\pm\,,
\end{equation}
which is the Pauli equation perturbed by the $O(\alpha^4)$ term;
note that our scaling assumptions (\ref{scaling}) imply
$eH^3\sim\alpha^3$, so the magnetic term cannot
be included in $O(\alpha^4)$. Similar arguments
(see also Section 33 in \cite{LL4},
as well as Theorem 6.8 in \cite{T})
reduce (\ref{DiracM3}) to (\ref{Pauli2}).
Thus, we have (formally)
shown that the energy levels of our model equation
(\ref{model1}) and those of the 2--dimensional
Dirac equation (\ref{DiracM3}) are related as
\begin{equation}
\label{relation}
\varepsilon_{\text{model}}=\varepsilon_{\text{Dirac}}+
O(\alpha^4)\,.
\end{equation}
Normally one subtracts the rest mass from $\varepsilon$ and
deals with $E:=\varepsilon-m\sim\alpha^2$,
so (\ref{relation}) means that the relative accuracy in the determination
of $E$ is $\sim\alpha^2$.

\newsection{Spectral properties of the model equation}

Throughout this section we assume that the electromagnetic
vector potential $A$ does not depend on $x^0$,
is smooth and vanishes at infinity.
In addition we assume
\begin{equation}
\label{restriction}
\|e\Phi\|_{L^\infty}<m\,,
\end{equation}
and that the first derivatives of
${\bold A}$ vanish at infinity.

Define the operator ${\cal P }$ formally by
\[
{\cal P}=\pmatrix P_2{}^2 & -P_2P_1 \\ -P_1P_2 & P_1{}^2\endpmatrix
=\pmatrix -P_2\\ P_1\endpmatrix\pmatrix -P_2 & \!\!P_1\endpmatrix
=\pmatrix -P_2 & \!\!P_1\endpmatrix^*\pmatrix -P_2 & \!\!P_1\endpmatrix.
\]
More precisely, we consider first the nonnegative symmetric operator
$C_0^\infty(\R^2)\to L^2(\R^2)$
given by the above
expression, and define ${\cal P }$ as the Friedrichs extension of the
latter; see Theorem 4.4.5 from \cite{D}.
Thus, ${\cal P}:D({\cal P})\to L^2(\R^2)$
is a nonnegative self--adjoint operator
defined on some
$D({\cal P})\supset{\cal S}(\R^2)\supset C_0^\infty(\R^2)$
(here ${\cal S}$ stands for the Schwartz class).

We have ${\cal A}={\cal P}+m^2I+e\Phi{\cal B}$, and the operator
${\cal A}:D({\cal A})\to L^2(\R^2)$, $D({\cal A})=D({\cal P})$,
is self--adjoint.
Furthermore,
condition (\ref{restriction}) means that $\|e\Phi{\cal B}\|<m^2$,
so the operator ${\cal A}$ is positive definite:
$a:=\inf\sigma({\cal A})>0$.
Therefore ${\cal A}^{-\frac12}$ is a bounded self--adjoint operator
in $L^2({\Bbb R}^2)$. We can now rewrite (\ref{spectral1}) as
\begin{equation}
\label{spectral2}
{\cal A}^{-\frac12}{\cal B}{\cal A}^{-\frac12}{\bold v}
=\frac1\varepsilon{\bold v}.
\end{equation}
The latter spectral problem is well--posed because
${\cal A}^{-\frac12}{\cal B}{\cal A}^{-\frac12}$ is a bounded self-adjoint
operator in $L^2(\R^2)$.

\begin{theorem}
\label{theorem3}
The essential spectrum of the operator
${\cal A}^{-\frac12}{\cal B}{\cal A}^{-\frac12}$ is the interval
$[-m^{-1},m^{-1}]$.
\end{theorem}

Theorem \ref{theorem3} says, in effect, that
for potentials which are
sufficiently weak, smooth and well behaved at infinity,
the basic structure
of the spectrum of our model equation
(\ref{model1}) is the same as that of
the Dirac equation (\ref{DiracM3}).

\begin{proofo} of Theorem \ref{theorem3}.

{\bf Part 1 of the proof.}
Let us prove
\begin{equation}
\label{part1}
\bigl[{-m}^{-1},m^{-1}\bigr]\subset
\sigma_\ess\bigl({\cal A}^{-\frac12}{\cal B}{\cal A}^{-\frac12}\bigr).
\end{equation}

Let $\varepsilon$ be an arbitrary real number such that $|\varepsilon|>m$.
Suppose we have a sequence $\{{\bold u}^{(n)}\}$ such that
\vspace{-5pt}
\begin{equation}
\label{sequence1}
{\bold u}^{(n)}\in C_0^\infty(\R^2),
\end{equation}
\vspace{-20pt}
\begin{equation}
\label{sequence2}
\big\|{\bold u}^{(n)}\big\|=1,
\end{equation}
\vspace{-20pt}
\begin{equation}
\label{sequence3}
\big\|
\big(
{\cal A}-\varepsilon{\cal B}
\big){\bold u}^{(n)}
\big\|\to0
\end{equation}
(a sequence of approximate eigenfunctions of problem (\ref{spectral1})).
Put ${\bold v}^{(n)}\!=\!
{\cal A}^{\frac12}{\bold u}^{(n)}\in
D({\cal A}^{\frac12})\subset L^2(\R^2)$.
We have
$\|{\bold v}^{(n)}\|\ge a^{\frac12}$ and
\[
\big\|\big(\varepsilon^{-1}\!
-{\cal A}^{-\frac12}{\cal B}{\cal A}^{-\frac12}
\big){\bold v}^{(n)}\big\|
=|\varepsilon|^{-1}
\big\|{\cal A}^{-\frac12}\big({\cal A}-\varepsilon{\cal B}
\big){\bold u}^{(n)}\big\|
\le|\varepsilon|^{-1}a^{-\frac12}\big\|\big({\cal A}-\varepsilon{\cal B}\big)
{\bold u}^{(n)}\big\|,
\]
so $\|(\varepsilon^{-1}\!-
{\cal A}^{-\frac12}{\cal B}{\cal A}^{-\frac12}){\bold v}^{(n)}
\|\to0$. This implies
$\varepsilon^{-1}\in
\sigma\bigl({\cal A}^{-\frac12}{\cal B}{\cal A}^{-\frac12}\bigr)$.
As $\varepsilon^{-1}$ is an arbitrary number in
$\bigl({-m}^{-1},m^{-1}\bigr)\setminus\{0\}$ we arrive at
(\ref{part1}). Thus, we have reduced the proof of (\ref{part1}) to
the construction of a sequence with properties
(\ref{sequence1})--(\ref{sequence3}).

Let us denote by ${\cal P}_0$ the operator ${\cal P}$
in the case ${\mathbf{A}}\equiv{\bold 0}$, and by
${\cal A}_0$ the operator ${\cal A}$
in the case $A\equiv0$; of course,
${\cal A}_0={\cal P}_0+m^2I$.

Suppose we have a sequence $\{{\bold u}^{(n)}\}$ with properties
(\ref{sequence1}), (\ref{sequence2}) and
\begin{equation}
\label{sequence4}
\big\|
\big(
{\cal A}_0-\varepsilon{\cal B}
\big){\bold u}^{(n)}
\big\|\to0\,.
\end{equation}
Let us modify this sequence by translating the functions, that is,
by replacing each ${\bold u}^{(n)}(x)$ by
${\bold u}^{(n)}(x-x^{(n)})$, where $x^{(n)}\in\R^2$.
Clearly, if the sequence of points $\{x^{(n)}\}$ is chosen
to tend to infinity sufficiently quickly, then
$\|({\cal A}-{\cal A}_0){\bold u}^{(n)}\|\to0$ and (\ref{sequence4})
will imply (\ref{sequence3}).
The proof of (\ref{part1}) has been reduced to
the construction of a sequence with properties
(\ref{sequence1}), (\ref{sequence2}), (\ref{sequence4}).

As functions of Schwartz class can be approximated by $C_0^\infty$
functions, we can relax the condition (\ref{sequence1}) by
replacing it with
\begin{equation}
\label{sequence5}
{\bold u}^{(n)}\in{\cal S}(\R^2).
\end{equation}

The differential operator ${\cal A}_0-\varepsilon{\cal B}$ has
constant coefficients, so it is natural to switch from the
${\bold u}^{(n)}(x)$ to their Fourier transforms
\[
\widehat{\bold u}^{(n)}(\xi):=\frac1{2\pi}\int
e^{-i\langle x,\xi\rangle}{\bold u}^{(n)}(x)\,d^2x\,,
\qquad\langle x,\xi\rangle=x^1\xi_1+x^2\xi_2.
\]
We now have to construct a sequence
$\{\widehat{\bold u}^{(n)}\}$ such that
\begin{equation}
\label{sequence6}
\widehat{\bold u}^{(n)}\in{\cal S}(\R^2),
\end{equation}
\vspace{-20pt}
\begin{equation}
\label{sequence7}
\big\|\widehat {\bold u}^{(n)}\big\|=1,
\end{equation}
\vspace{-20pt}
\begin{equation}
\label{sequence8}
\big\|
\big(
\widehat{\cal A}_0-\varepsilon{\cal B}
\big)\widehat{\bold u}^{(n)}
\big\|\to0\,,
\end{equation}
where
\[
\widehat{\cal A}_0-\varepsilon{\cal B}\equiv
\widehat{\cal A}_0(\xi)-\varepsilon{\cal B}=
\pmatrix
\xi_2^2+m^2& -\xi_2\xi_1\pm i\varepsilon m\\
-\xi_1\xi_2\mp i\varepsilon m&\xi_1^2+m^2\endpmatrix
\]
is the (full) symbol of ${\cal A}_0-\varepsilon{\cal B}$. We have
$
\det(\widehat{\cal A}_0(\xi)-\varepsilon{\cal B})=
m^2(\xi_1^2+\xi_2^2+m^2-\varepsilon^2)
$
and $|\varepsilon|>m$, so
we can choose an $\eta$ to give
$\det(\widehat{\cal A}_0(\eta)-\varepsilon{\cal B})=0$. Let
$\widehat{\bold u}$ be a normalised (constant) vector
in the null space of
$\widehat{\cal A}_0(\eta)-\varepsilon{\cal B}$, and let
$\{\phi^{(n)}\}$ be a sequence of scalar functions such that
$\phi^{(n)}\in C_0^\infty(\R^2)$, $\|\phi^{(n)}\|=1$,
and $\supp\phi^{(n)}\to\{\eta\}$. It is easy to see
that the vector functions
$\widehat{\bold u}^{(n)}(\xi):=\widehat{\bold u}\,\phi^{(n)}(\xi)$
have the required properties
(\ref{sequence6})--(\ref{sequence8}).

{\bf Part 2 of the proof.}
Let us prove
$\sigma_\ess({\cal A}^{-\frac12}{\cal B}{\cal A}^{-\frac12})
\subset[-m^{-1},m^{-1}]$.

Since $\|{\cal B}\|=m$ it
is sufficient to show
$\sigma_\ess({\cal A})\subset[m^2,+\infty)$.
However, ${\cal A}={\cal P}+m^2I+e\Phi{\cal B}$,
so, in turn, it is sufficient to show
\begin{equation}
\label{part2}
\sigma_\ess({\cal P}+\Psi{\cal B})\subset[0,+\infty)\,,
\end{equation}
where $\Psi=e\Phi$ is an arbitrary smooth real valued
function vanishing at infinity.

\emph{Claim: for any $\Psi\in C^\infty_0(\R^2)$ we have}
\begin{equation}
\label{essential1}
\sigma_\ess\bigl({\cal P}_0+I+\Psi{\cal B}\bigr)\subset[0,+\infty)\,.
\end{equation}

Indeed, let us define the operators ${\cal Q}_0:={\cal P}_0+I$,
${\cal R}:={\cal Q}_0^{-\frac12}
\Psi{\cal B}{\cal Q}_0^{-\frac12}$.
The operator ${\cal Q}_0$ is a positive definite
differential operator with constant coefficients, and its symbol is
\[
{\cal Q}_0(\xi)=\pmatrix \xi_2^2+1& -\xi_2\xi_1\\ -\xi_1\xi_2&
\xi_1^2+1\endpmatrix.
\]
Therefore ${\cal Q}_0^{-\frac12}$ is a
pseudodifferential operator with symbol
\[
{\cal Q}_0^{-\frac12}(\xi)=
\pmatrix 1& 0\\ 0& 1\endpmatrix-
\frac1{\Xi(1\!+\!\Xi)}
\pmatrix \xi_2^2& -\xi_2\xi_1\\ -\xi_1\xi_2& \xi_1^2\endpmatrix,
\quad\Xi=\bigl(1+\xi_1^2+\xi_2^2\bigr)^{\frac12}.
\]
The principal symbol of the operator ${\cal Q}_0^{-\frac12}$ is
\[
M(\xi)=
\pmatrix 1& 0\\ 0& 1\endpmatrix-
\frac1{\xi_1^2+\xi_2^2}
\pmatrix \xi_2^2& -\xi_2\xi_1\\ -\xi_1\xi_2& \xi_1^2\endpmatrix,
\]
so ${\cal Q}_0^{-\frac12}$
is a pseudodifferential operator of order 0.
It follows that ${\cal R}$ is a
pseudo\-differential operator of order 0.
The principal symbol of ${\cal R}$ is
\[
\pm\,im\,M(\xi)
\pmatrix \!\!0& \!\!-\Psi(x) \\\Psi(x)\!\!& 0\!\!\endpmatrix
M(\xi)\,=\,0\,,
\]
so ${\cal R}$ is in fact a pseudodifferential operator of
order -1. Now, the Schwartz kernel
$K_{\cal R}(x,y)$ of the operator ${\cal R}$ is given by
the oscillatory integral
\[
\pm\,\frac{im}{(2\pi)^2}\int\!
{\cal Q}_0^{-\frac12}(\xi)
\,e^{i\langle x-z,\xi\rangle}\!
\pmatrix \!0& \!-\Psi(z) \\ \Psi(z)\!& 0\!\endpmatrix\!
{\cal Q}_0^{-\frac12}(\eta)
\,e^{i\langle z-y,\eta\rangle}d^2\xi\,d^2z\,d^2\eta\ .
\]
Let $\Omega\subset\R^2$ be a bounded open set with
$\supp\Psi\subset\Omega$.
A standard calculation shows that $K_{\cal R}(x,y)$ is smooth and
rapidly decreasing in both $x$ and $y$ outside
$\Omega\!\times\!\Omega$
(n.b.~this calculation relies on the fact that the matrix function
${\cal Q}_0^{-\frac12}(\xi)$ is smooth on the whole of $\R^2$,
including the point $\xi=0$).
Since $K_{\cal R}(x,y)$ is the integral kernel
of a pseudodifferential operator of order -1 inside $\Omega$,
it follows that the operator ${\cal R}$ is compact. Hence
the operator $I+{\cal R}$ is nonnegative on a subspace
of finite co-dimension. Consequently, the quadratic form
\[
\int{\bold v}^*(I+{\cal R})
{\bold v}\,d^2x\,,\qquad{\bold v}\in L^2(\R^2)\,,
\]
is nonnegative on a subspace of finite co-dimension. Equivalently,
\[
\int\big({\cal Q}_0^{\frac12}{\bold u}\big)^*
(I+{\cal R}){\cal Q}_0^{\frac12}
{\bold u}\,d^2x\,,\qquad{\bold u}\in D\big({\cal
P}_0^{\frac12}\big)\,,
\]
is nonnegative on a subspace of finite co-dimension. But the latter
is the quadratic form associated with the operator
${\cal P}_0+I+\Psi{\cal B}$. Formula (\ref{essential1})
now follows.

Let us now remove the restriction $\Psi\in C^\infty_0(\R^2)$ and
show that (\ref{essential1}) still holds. Given an arbitrary
$\epsilon>0$ we can decompose $\Psi$ as $\Psi=\Psi_0+\Psi_1$ where
$\Psi_0\in C^\infty_0$ and $\|\Psi_1\|_{L^\infty}\le m^{-1}\epsilon$.
We have
\[
{\cal P}_0+I+\Psi{\cal B}=({\cal P}_0+I+\Psi_0{\cal B})+\Psi_1{\cal B}.
\]
According to (\ref{essential1})
$\sigma_\ess({\cal P}_0+I+\Psi_0{\cal B})\subset[0,+\infty)$, whereas
$\Psi_1{\cal B}$ is a bounded operator
with $\|\Psi_1{\cal B}\|\le\epsilon$. This implies
$\sigma_\ess({\cal P}_0+I+\Psi{\cal B})\subset[-\epsilon,+\infty)$.
As $\epsilon>0$ is arbitrary we arrive at (\ref{essential1})
for general $\Psi$.

Now, given an arbitrary $\delta\in(0,1)$ we have
\[
{\cal P}_0+\Psi{\cal B}\ge
\delta{\cal P}_0+\Psi{\cal B}
=\delta({\cal P}_0+I+\delta^{-1}\Psi{\cal B})-\delta I.
\]
According to (\ref{essential1})
$\sigma_\ess({\cal P}_0+I+\delta^{-1}\Psi{\cal B})\subset[0,+\infty)$, so
$\sigma_\ess({\cal P}_0+\Psi{\cal B})\subset[-\delta,+\infty)$.
As $\delta\in(0,1)$ is arbitrary we conclude that
\begin{equation}
\label{essential2}
\sigma_\ess\bigl({\cal P}_0+\Psi{\cal B}\bigr)\subset[0,+\infty)\,.
\end{equation}

Thus, we have proved (\ref{part2}) in the case ${\bold A}\equiv{\bold 0}$.
Let us now remove this restriction.
For any $\nu>0$ and any smooth vector function ${\bold u}$
we have pointwise
\[
\big|P_1u^2-P_2u^1\big|^2=
\frac{\nu}{1+\nu}\big|\nabla_1u^2-\nabla_2u^1\big|^2
-\nu e^2\big|A_1u^2-A_2u^1\big|^2
+\big|\widetilde P_1u^2-\widetilde P_2u^1\big|^2,
\]
where
$\,\widetilde P_\varkappa:=
i(1+\nu)^{-\frac12}\nabla_\varkappa
-e(1+\nu)^{\frac12}A_\varkappa\,$, $\,\varkappa=1,2$.
This implies
\[
\int{\bold u}^*{\cal P}{\bold u}\,d^2x\,\ge\,
\frac{\nu}{1+\nu}\int{\bold u}^*{\cal P}_0{\bold u}\,d^2x\,
-\,C\,\nu\,\|{\bold u}\|^2\,,
\qquad {\bold u}\in C_0^\infty(\R^2),
\]
with $\,C\,=\,e^2\,\big\|{A_1}^2+{A_2}^2\big\|_{L^\infty}\,$.
Consequently,
\[
\int{\bold u}^*\bigl({\cal P}+\Psi{\cal B}\bigr){\bold u}\,d^2x\,\ge
\,\frac{\nu}{1+\nu}
\int{\bold u}^*\bigl({\cal P}_0+
\widetilde\Psi{\cal B}\bigr){\bold u}\,d^2x\,
-\,C\,\nu\,\|{\bold u}\|^2\,,
\qquad {\bold u}\in C_0^\infty(\R^2)\,,
\]
with $\widetilde\Psi=\nu^{-1}(1+\nu)\Psi$.
The latter formula and (\ref{essential2}) imply
$\sigma_\ess({\cal P}+\Psi{\cal B})
\subset[-C\nu,\,+\infty)$.
As $\nu>0$ is arbitrary we arrive at (\ref{part2}).
\end{proofo}

Comprehensive analysis of the discrete spectrum of
the operator
${\cal A}^{-\frac12}{\cal B}{\cal A}^{-\frac12}$
is a non--trivial task which lies
outside the scope of this paper. We shall, however, briefly
deal with the most basic situation.

Suppose $\Phi\in{\cal S}(\R^2)$ and satisfies (\ref{restriction}),
and suppose ${\bold A}\equiv0$. Let $\varepsilon^{-1}$,
$|\varepsilon|<m$, be an eigenvalue of (\ref{spectral2}).
Then, using arguments similar to those in the second part of the
proof of Theorem \ref{theorem3}, one can show that
${\bold u}\in{\cal S}(\R^2)$. Therefore, in studying
the discrete spectrum
we can work with (\ref{spectral1}) or (\ref{unitary})
rather than with (\ref{spectral2}).

Suppose now $\Phi$ is radially symmetric.
Let us introduce polar coordinates
$x^1=r\cos\vartheta$, $x^2=r\sin\vartheta$, and
expand the $\varphi_\pm$ as
\begin{equation}
\label{expansion}
\varphi_\pm=\sum_{k\in\Z}e^{i(k\mp1)\vartheta}
\varphi_\pm^{(k)}(r)\,.
\end{equation}
Substituting these expansions into (\ref{unitary}) we see that
the latter separates into systems of ordinary
differential equations
\begin{equation}
\label{Radial1}
\pmatrix
H_{k-1}^\pm& G_k\\
G_{-k}& H_{k+1}^\mp
\endpmatrix
\pmatrix\varphi_+^{(k)}\\\varphi_-^{(k)}\endpmatrix=0\,,
\end{equation}
where $k$ runs through $\Z$, and
\[
H_l^\pm:=m-
\frac1{2m}\left(
\frac1r\frac d{dr}r\frac d{dr}-\frac{l^2}{r^2}\right)
\mp(\varepsilon-e\Phi)\,,
\]
\[
G_l:=\frac1{2m}\left(
\frac1r\frac d{dr}r\frac d{dr}+\frac{2l}r\frac d{dr}+
\frac{l^2-1}{r^2}\right).
\]
The orthogonal transformation
\[
\pmatrix\varphi_+^{(k)}\\\varphi_-^{(k)}\endpmatrix=
\frac1{\sqrt2}
\pmatrix1& 1\\1&-1\endpmatrix
\pmatrix f\\ g\endpmatrix
\]
turns (\ref{Radial1}) into
\begin{equation}
\label{Radial2}
\frac{k}rg'+\frac{k}{r^2}g-\frac{k^2}{r^2}f
-m^2f\pm m(\varepsilon-e\Phi)g=0\,,
\end{equation}
\vspace{-12pt}
\begin{equation}
\label{Radial3}
g''+\frac1rg'-\frac1{r^2}g
-\frac{k}rf'+\frac{k}{r^2}f
-m^2g\pm m(\varepsilon-e\Phi)f=0\,,
\end{equation}
where the prime stands for the derivative in $r$.
Resolving (\ref{Radial2}) with respect to $f$ and substituting
the resulting expression into (\ref{Radial3}), we reduce our
problem to a single second order equation
\begin{multline}
\label{Radial4}
g''+\frac{3k^2+m^2r^2}{(k^2+m^2r^2)r}\;g'\\
+\left(\frac{k^2-m^2r^2}{(k^2+m^2r^2)r^2}
-\frac{k^2}{r^2}\pm\frac{2km(\varepsilon-e\Phi)}{k^2+m^2r^2}
\pm\frac{ke\Phi'}{mr}+(\varepsilon-e\Phi)^2-m^2\right)g=0\,.
\end{multline}

Let us compare our model equation (\ref{Radial4})
with the Klein--Gordon equation
\begin{equation}
\label{KG}
g''+\frac1rg'+
\left(-\frac{n^2}{r^2}+(\varepsilon-e\Phi)^2-m^2\right)g=0\,,
\end{equation}
$n\in\Z$. In view of (\ref{unitary}), (\ref{expansion})
it is natural to compare the two equations taking
$|n|=|k\mp1|$ when we are looking for the bound states
of the electron ($\varepsilon\approx+m$), and
$|n|=|k\pm1|$ when we are looking for the bound states
of the positron ($\varepsilon\approx-m$).
Clearly, for $k=0$ (\ref{Radial4}) and (\ref{KG}) coincide.
For $k\ne0$ these equations differ and we shall compare
their spectra asymptotically, assuming
that $e\Phi$ is of the form
$\alpha^2\Psi(\alpha r)$, $\Psi\in{\cal S}(\R^2)$, $\alpha\to+0$;
this is a particular case of the situation (\ref{scaling}).
Asymptotic analysis gives
$\varepsilon_{\text{model}}=
\varepsilon_{\text{KG}}+O(\alpha^4)$,
which in turn implies
$\varepsilon_{\text{model}}=
\varepsilon_{\text{Dirac}}+O(\alpha^4)$.

We hope to justify a version of (\ref{relation}) in the
case of more general $\Phi$ and ${\bold A}$ by means of
asymptotic perturbation techniques (viz. Chapter VIII of \cite{K}).
Here the technical difficulty is that
in contrast to the Dirac operator which can be viewed as an analytic
perturbation of the Pauli operator (see Chapter 6 of \cite{T}
for more details) the perturbation in (\ref{Pauli2}) is not analytic.

\newsection{Conclusion}

The results of Sections 6 and 7 seem to indicate that
the basic effects attributed to spinors can be explained
(at least in Minkowski 3--space)
using simple tensor models. This observation
may be useful in relation to attempts
at modelling the electron as a
soliton--like solution of some nonlinear system
of partial differential equations. The usual
approach, see, e.g., \cite{W} and \cite{EGS},
involves the so--called Maxwell--Dirac equation.
In our view, it might make sense looking also at other
nonlinear systems which do not necessarily have
spinors occurring explicitly but may still produce spinor
effects.
With this goal in mind,
let us compare our model equation
(\ref{model}) with the Maxwell system.

Define the operator of exterior differentiation
$d$ mapping vectors to antisymmetric tensors as
$(du)_{\mu\nu}:=i\nabla_\mu u_\nu-i\nabla_\nu u_\mu$,
and its dual $\delta$ as
$(\delta T)^\nu:=i\nabla_\mu T^{\mu\nu}$;
we also define the action of $\delta$ on vectors
as $\delta u:=i\nabla_\mu u^\mu$.
By analogy, define
$(d_Au)_{\mu\nu}:=P_\mu u_\nu-P_\nu u_\mu$,
$(\delta_A T)^\nu:=P_\mu T^{\mu\nu}$.
Now, squaring (\ref{model}) gives
$\,{\operatorname{curl}_A}^2u=m^2u\,$. But
$\,{\operatorname{curl}_A}^2=\delta_Ad_A$, so
our model equation (\ref{model}) becomes
\begin{equation}
\label{model3}
\delta_Ad_Au=m^2u\,.
\end{equation}
According to formula (30.2) of \cite{LL2}
the Maxwell system can be written as
\begin{equation}
\label{Maxwell}
\delta d A=-4\pi j\,,\qquad\delta A=0\,,
\end{equation}
where $j$ is the (given) current. The similarity between
(\ref{model3}) and (\ref{Maxwell}) is remarkable, and
we hope to build further mathematical models on the basis of
this similarity.

\address{School of Mathematical Sciences\\
University of Sussex\\
Falmer\\
Brighton BN1 9QH\\
United Kingdom}
\address{School of Mathematical Sciences\\
University of Sussex\\
Falmer\\
Brighton BN1 9QH\\
United Kingdom}

\received{Date inserted by the Editor}

\end{document}